\documentclass{jltp}

\usepackage{graphicx}

\title{Anisotropic Aerogels for Studying Superfluid $^3$He}

\author{J. Pollanen, S. Blinstein, H. Choi, J.P. Davis,\\T.M. Lippman, L.B. Lurio$^*$, and W.P. Halperin}

\address{Department of Physics and Astronomy, Northwestern University, Evanston, IL\\ 60208, USA\\$^*$Department of Physics, Northern Illinois University, DeKalb, IL 60115, USA}

\runninghead{J. Pollanen \em{et al.}}{Anisotropic Aerogels for Studying Superfluid $^3$He}

\begin{document}

\maketitle

\begin{abstract}
It may be possible to stabilize new superfluid phases of $^{3}$He with anisotropic silica aerogels.  We discuss two methods that introduce anisotropy in the aerogel on length scales relevant to superfluid $^{3}$He.  First, anisotropy can be induced with uniaxial strain.  A second method generates anisotropy during the growth and drying stages.  We have grown cylindrical $\sim$98$\%$ aerogels with anisotropy indicated by preferential radial shrinkage after supercritical drying and find that this shrinkage correlates with small angle x-ray scattering (SAXS).  The growth-induced anisotropy was found to be $\sim$90$^\circ$ out of phase relative to that induced by strain.  This has implications for the possible stabilization of superfluid phases with specific symmetry.

\end{abstract}
\section{INTRODUCTION}
Recently, there has been interest in stabilizing new phases of superfluid $^3$He not found in bulk.  This might be possible for $^3$He within a silica aerogel that is anisotropic on the length scale of the superfluid coherence length.  Aerogel consists of silica strands with a diameter of $\sim$30 {\AA} and average separation, or correlation length, $\xi_a \cong$ $300-1000$ {\AA}.  It is thought that anisotropy in the quasiparticle scattering plays an important role in determining which superfluid $^3$He phases are energetically stable\cite{Thu98,Vic05,Aoy05}.  For instance, it is predicted that isotropic quasiparticle scattering from aerogel tends to stabilize the isotropic {\em{B}}-phase relative to the anisotropic {\em{A}}-phase\cite{Thu98}.  Conversely, Vicente {\em{et al.}}\cite{Vic05} proposed using uniaxially strained aerogels to study the effect of global anisotropy on the stabilization of the anisotropic {\em{A}}-phase.  In addition, Aoyama and Ikeda\cite{Aoy05} have predicted that anisotropy introduced by strain in the form of uniaxial {\em compression} or {\em elongation} could lead to modifications of the phase diagram for $^3$He in aerogel.  These authors find that an equal spin pairing state is stabilized for either type of strain and indicate that for $^3$He in a uniaxially {\em elongated} aerogel a polar pairing state could be realized.  In preliminary experimental work, Davis {\em et al.}\cite{Dav06} observed two stable phases in $^3$He in aerogel which they attributed to anisotropy in their aerogel sample.

We have previously performed small angle x-ray scattering (SAXS) on uniaxially compressed aerogels and found\cite{Pol06} that strain introduced anisotropy on the order of $\sim$$ 100$ {\AA}, and that this anisotropy scaled with the applied strain.  To explore the possibility of introducing anisotropy different from that produced by compression, we have performed SAXS on a series of aerogels which exhibited preferential radial shrinkage after supercritical drying.  We found that shrinkage introduces anisotropy on the length scale of the aerogel correlation length, $\xi_{a}$, and that the direction of this growth-induced anisotropy was approximately perpendicular to that produced by uniaxial compression.
\section{EXPERIMENT}
SAXS studies of a series of aerogel samples ($\sim$$97$\%$-98$$\%$ porosity) were performed at Sector 8 of the Advanced Photon Source (APS) at Argonne National Laboratory, using a photon energy of 7.5 keV.  The samples were grown at Northwestern University by the ``one-step'' sol-gel method\cite{Fri85}.  It should be noted that the samples from our previous work\cite{Pol06} were grown via the ``two-step'' method\cite{Till92}.  We have observed that aerogel samples exhibit increased shrinkage when the pH is increased during the gelation process and when the heating rate in the autoclave is increased during supercritical drying.  By utilizing these two ``control'' parameters cylindrical samples were grown exhibiting radial shrinkage in the range from $0\%$ to $20\%$ and nominally no vertical shrinkage.  The samples were grown in borosilicate glass tubes and the amount of shrinkage was determined with precision calipers.  The cylindrical samples were cut to specific lengths with aspect ratios of $\sim$1.  We also performed SAXS on nominally unshrunken samples that were compressed along the cylinder axis in order to compare the two types of anisotropy.

The x-ray beam illuminated the center of a sample, which was oriented such that the cylinder axis was perpendicular to the beam.  For each sample we obtained the scattered x-ray intensity, {\em I}({\em q}), as a function of the momentum transfer, {\em q}.  The momentum transfer is defined as $q = (4\pi/\lambda) \sin(\theta/2)$, where $\lambda$ is the wavelength of the incident x-ray and $\theta$ is the angle between the incident and scattered x-ray wavevectors.  We binned {\em I}({\em q}) in $\sim$10$^\circ$ increments of the azimuthal angle, $\phi$, defined in the plane of the CCD camera.  For reference, $\phi = 90^\circ$ is parallel to the cylinder axis.

\section{RESULTS AND DISCUSSION}
\begin{figure}
\centerline{\includegraphics[width=0.9\textwidth]{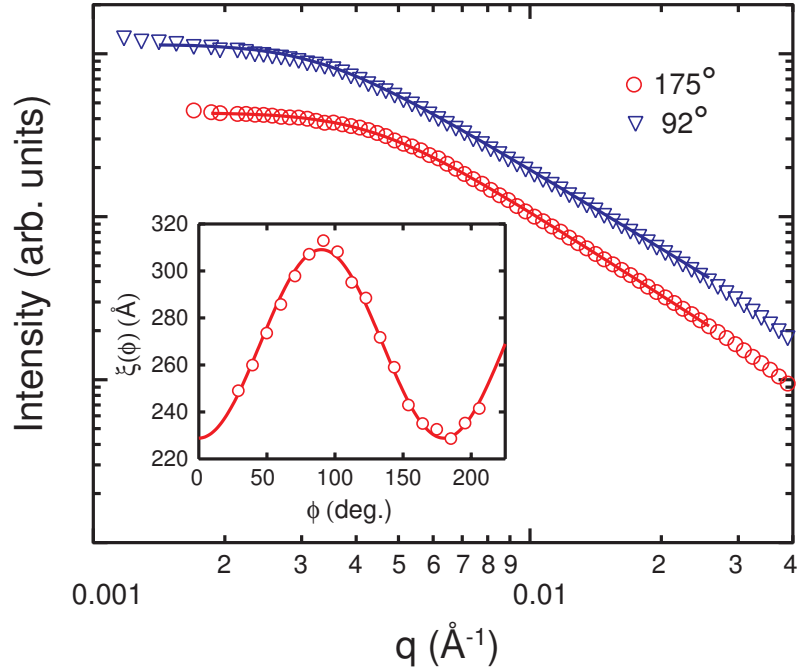}}
\caption{\label{fig1}(Color online) Scattering curves fit with Eq. (1) ($\phi = 92^\circ, 175^\circ$) for an aerogel exhibiting $4.6\%$ radial shrinkage.  The inset depicts $\xi(\phi)$.  The two curves have been offset vertically for clarity, otherwise the data would coincide at high $q$.}  
\end{figure}

We analyzed the scattering curves by fitting {\em I}({\em q}) with the following phenomenological scattering function,
\begin{equation}
I(q)={{C\xi^{d}}\over{(1+q^{4}\xi^{4})^{d/4}}}{{(1+q^{2}\xi^{2})^{1/2}}\over{q\xi}}\sin[(d-1)\tan^{-1}(q\xi)],
\end{equation}
where $C$, $d$, and $\xi$ are fit parameters.  Eq. (1) is a modified version of a structure factor for fractally correlated materials that have upper and lower length scale cut-offs\cite{Fre86}.  The length scale $\xi$ is usually associated with the aerogel correlation length, the parameter $d$ is identified with the fractal dimension of the aerogel, and $C$ is a scaling factor.  This form matched the data well.  Fig.~\ref{fig1} presents scattering curves for a sample shrunken by 4.6$\%$ ($\phi = 92^\circ$, $\phi = 175^\circ$).  The solid lines in Fig.~\ref{fig1} are fits using Eq. (1).

\begin{figure}
\centerline{\includegraphics[width=1\textwidth]{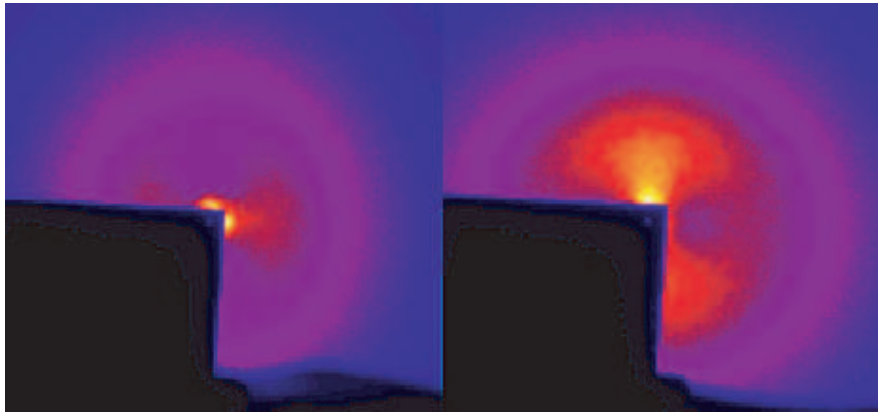}}
\caption{\label{fig2}(Color online) Scattered x-ray intensities for two $\sim$98$\%$ aerogel samples depicting the anisotropy produced by uniaxial strain (left) and radial shrinkage (right).  In each case the sample is oriented with its cylinder axis vertical relative to the images.  The dark rectangle is a beamstop inserted to block the unscattered transmitted beam.} \end{figure}

For each sample we plotted $\xi$ vs. $\phi$, and found it to vary as
\begin{equation}
\xi(\phi)=\xi_{0}-\xi_{1}\cos(2\phi),
\end{equation}
where $\xi_{0}$ and $\xi_{1}$ are parameters fit to the data.  This result, for the sample shrunken by $4.6\%$, is depicted in the inset of Fig.~\ref{fig1}.  Sinusoidal dependence of $\xi(\phi)$ for all samples clearly demonstrates that intrinsic anisotropy is present.
Moreover, this anisotropy is $\sim$90$^\circ$ out of phase with that induced by uniaxial strain\cite{Pol06}.  When the aerogel is compressed in the axial direction the scattering moves to smaller $q$ in the direction perpendicular to the cylinder axis.  This indicates that the axial length scales are compressed relative to the radial length scales.  The difference between strain and shrinkage is depicted in Fig.~\ref{fig2}, which presents false color images of the scattered x-ray intensity from two $\sim$98$\%$ aerogels as captured by the CCD camera.  The left panel is an image of an aerogel compressed by $12\%$, while the right panel is of a sample that exhibited $12.7\%$ radial shrinkage.  This 90$^\circ$ phase difference is consistent with our previous work\cite{Pol06}.
\begin{figure}
\centerline{\includegraphics[width=0.9\textwidth]{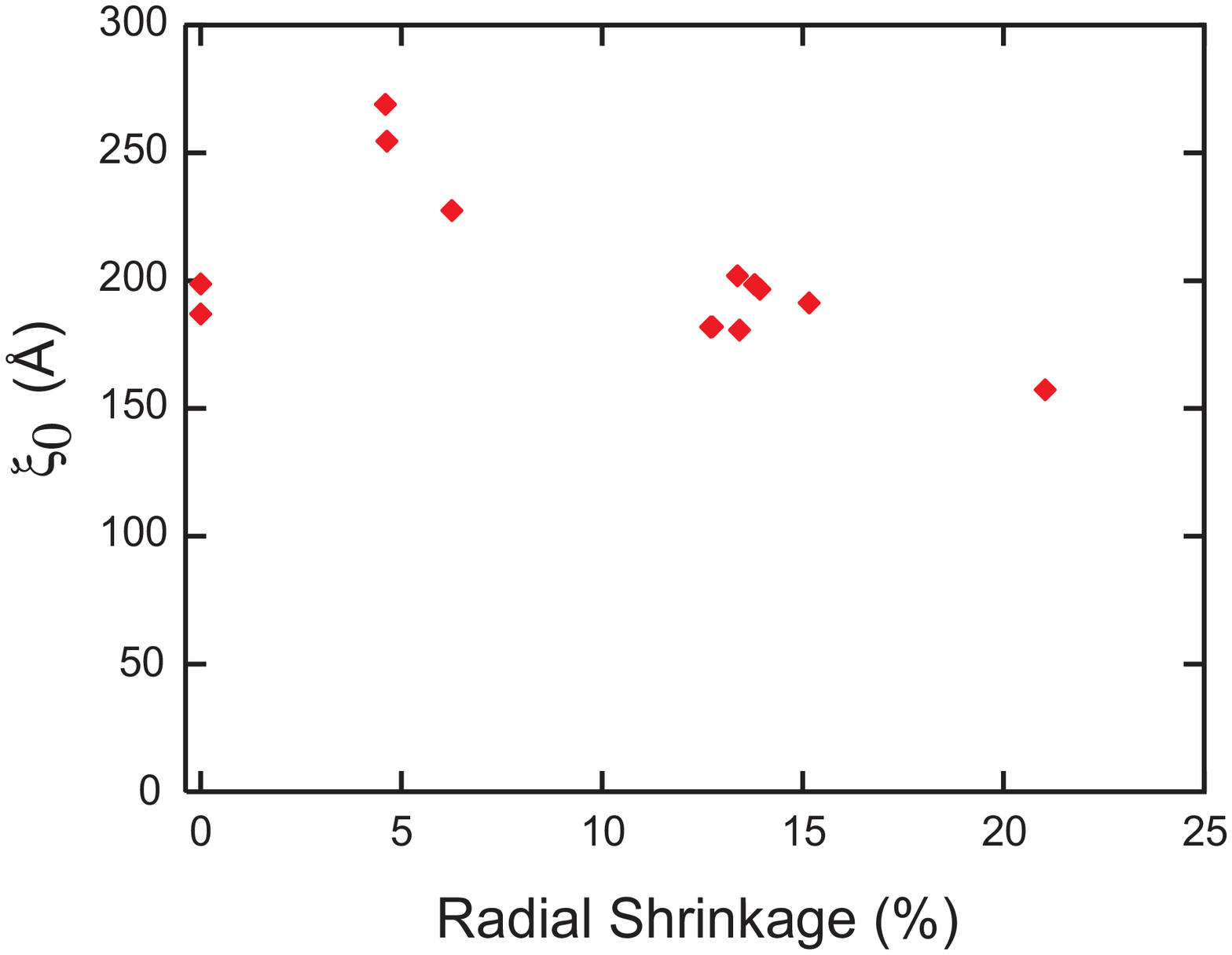}}
\caption{\label{fig3}(Color online) The fit parameter $\xi_{0}$ as a function of radial shrinkage.}  
\end{figure}
\begin{figure}
\centerline{\includegraphics[width=0.9\textwidth]{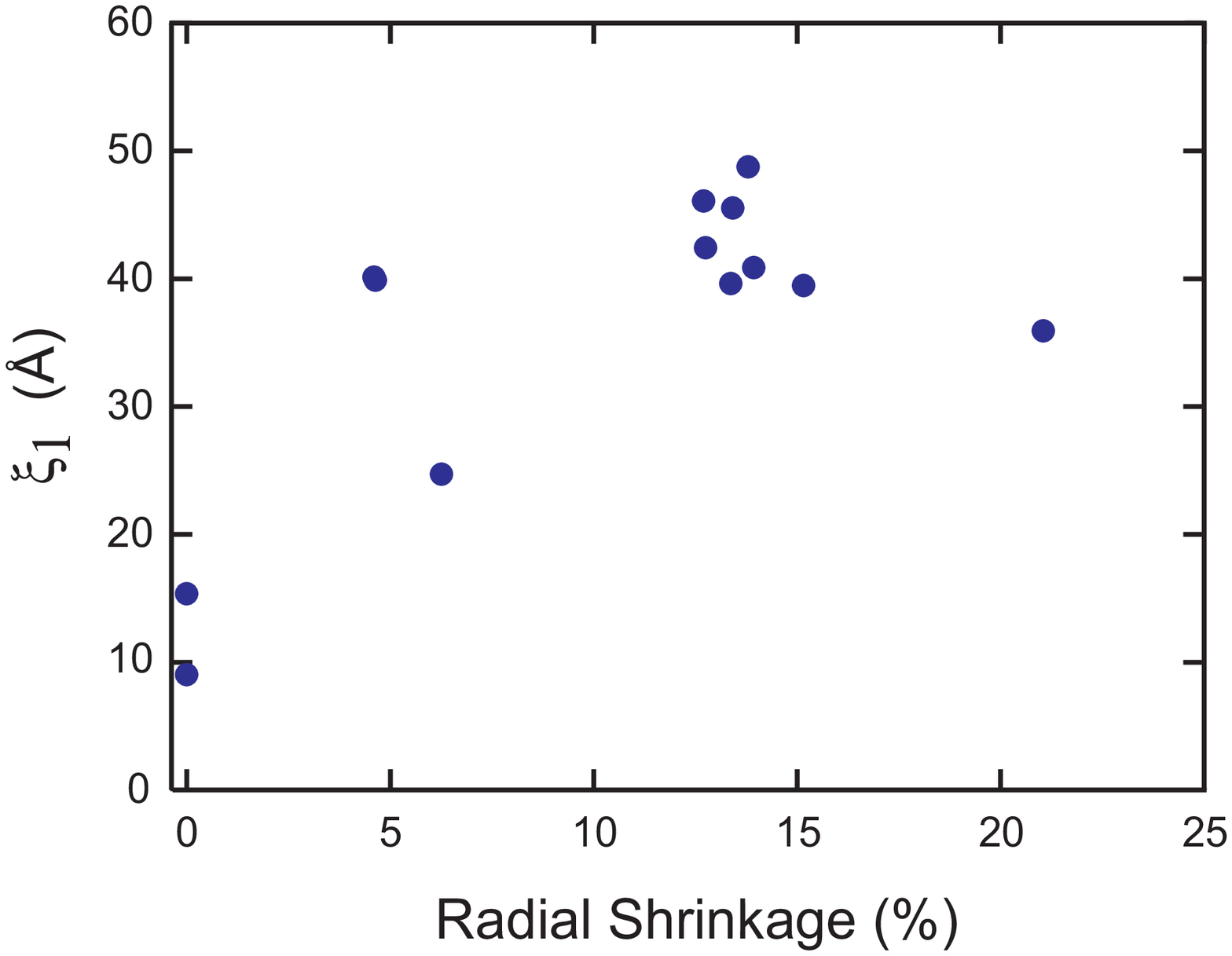}}
\caption{\label{fig4}(Color online) The fit parameter $\xi_{1}$ as a function of radial shrinkage.}  
\end{figure}
The coefficient $\xi_{0}$ was found to be roughly constant as a function of shrinkage with an average value of $\sim$$200$ {\AA}.  These results are presented in Fig.~\ref{fig3}.  The amplitude, $\xi_{1}$, of the anisotropy appears to increase with shrinkage.  In Fig.~\ref{fig4} we present these results demonstrating that anisotropy can be introduced into the aerogel during the growth and drying stages.  Concerning the dispersion of the data in Fig.~\ref{fig4}, it should be noted that SAXS is a local probe whereas the radial shrinkage is determined from the macroscopic geometry of the sample.  However, the small anisotropy of the unshrunken samples is prominent.

In conclusion, from our SAXS experiments we have demonstrated two methods of introducing anisotropy on the length scales relevant to superfluid $^{3}$He.  Anisotropy can result during the growth and drying stages.  Also, uniaxial strain can be used to induce anisotropy that is $\sim$90$^\circ$ out of phase with the growth-induced anisotropy.  It will be particularly interesting to conduct low temperature experiments to see how these types of anisotropy affect the phase diagram of superfluid $^3$He.  This is especially true in light of theoretical predictions\cite{Aoy05} suggesting  that the stability of various pairing states is dependent on the nature of the uniaxial anisotropy.

\section*{ACKNOWLEDGMENTS}
We would like to thank J.A. Sauls for valuable theoretical insights.  Use of the Advanced Photon Source was supported by the U. S. Department of Energy, Office of Science, Office of Basic Energy Sciences, under Contract No. W-31-109-ENG-38.  We also acknowledge the support of the National Science Foundation, DMR-0244099.

\end{document}